\documentclass[12pt]{elsart}
\usepackage{epsfig}

\begin{document}
\vspace{-1.0cm}
\begin{flushleft}
{\normalsize MIT-CTP-2978} \hfill\\
{\normalsize May 2000} 
\end{flushleft}
\vspace{1.0cm}
\begin{frontmatter}
  \title{\bf Perturbative renormalization of the first two moments of 
   non-singlet quark distributions with overlap fermions}
  
   \author{Stefano Capitani}
  
   \address{Center for Theoretical Physics, Laboratory for Nuclear 
   Science \\ Massachusetts Institute of Technology \\ 77 Massachusetts Ave., 
   Cambridge, MA 02139, USA \\ \ \\ {\sf E-mail: stefano@mitlns.mit.edu} }

  \date{ }

\begin{abstract}
Using the overlap-Dirac operator proposed by Neuberger, we have computed in
lattice QCD the one-loop renormalization factors of ten operators which 
measure the lowest two moments of unpolarized and polarized non-singlet 
quark distributions. These factors are necessary to extract physical numbers 
from Monte Carlo simulations made with overlap fermions.

An exact chiral symmetry is maintained in all our results, and the 
renormalization constants of corresponding unpolarized and polarized 
operators which differ by a $\gamma_5$ matrix have the same value.
We have considered two lattice representations for each continuum operator. 
The computations have been carried out using the symbolic language FORM, 
in a general covariant gauge. In some simple cases they have also been 
checked by hand. 
\end{abstract}
\end{frontmatter}

\newpage

\section{Introduction}

In recent years remarkable progress has been made towards formulations of 
lattice fermions which have no doublers and possess an exact chiral symmetry 
without giving up desirable features like flavor symmetry, locality, 
unitarity or gauge-invariance. The Ginsparg-Wilson relation~\cite{gw}
\begin{equation}
\gamma_5 D + D \gamma_5 = a \frac{1}{\rho} D \gamma_5 D
\label{eq:gw}
\end{equation}
has been recognized~\cite{hln} as the fundamental condition in this context, 
and a Dirac operator $D$ which satisfies this relation can indeed describe
chiral fermions on the lattice. One of the possible solutions of the 
Ginsparg-Wilson relation has been found by Neuberger~\cite{n2} starting from 
the overlap formalism~\cite{overlap}. In the massless case the overlap-Dirac 
operator is
\begin{equation}
D_N = \frac{1}{a} \rho \, \Big[ 1+ \frac{X}{\sqrt{X^\dagger X}} \Big] ,
\label{eq:dn}
\end{equation}
where
\begin{equation}
X=D_W -\frac{1}{a} \rho
\end{equation}
in terms of the usual Wilson-Dirac operator
\begin{equation}
D_W = \frac{1}{2} \Big[ \gamma_\mu (\nabla^\star_\mu + \nabla_\mu)
- a r \nabla^\star_\mu \nabla_\mu \Big] ,
\end{equation}
\begin{equation}
\nabla_\mu \psi (x) = 
\frac{1}{a} \Big[ U(x,\mu)\psi(x+a\hat{\mu}) - \psi (x) \Big] .
\end{equation}
In the range $0 <\rho <2r$ the right spectrum of massless fermions is 
obtained~\cite{n}. For a quark of bare mass $m_0$ the overlap-Dirac operator 
becomes
\begin{equation}
\Big(1-\frac{1}{2\rho}am_0\Big) D_N + m_0 .
\end{equation}
Since additive mass renormalization is forbidden, one avoids altogether 
a source of systematic errors that is always present with Wilson 
fermions~\cite{h}.

L\"uscher has shown~\cite{l2} that a fermion obeying the Ginsparg-Wilson 
relation possesses an exact chiral symmetry of the general form 
\begin{equation}
\delta \psi = \epsilon \cdot \gamma_5 \Big(1-\frac{c}{\rho} \, 
a D \Big) \psi, \quad
\delta \bar{\psi} = \epsilon \cdot \bar{\psi} \Big(1-\frac{1-c}{\rho} \, 
a D \Big) \gamma_5.
\label{eq:symmetry}
\end{equation}
A fully legitimate form of global chiral symmetry is thus preserved for 
finite lattice spacing. The expected value of the global anomaly~\cite{l2} 
(which comes from the non-invariance of the fermionic integration measure 
under the transformations above), the Atiyah-Singer index theorem on the 
lattice~\cite{hln} and the chiral Ward identities are then fully attained 
before taking the continuum limit. Thus, the central point in this framework 
is not to insist on the lattice form of the canonical chiral transformations 
(in fact actions satisfying Eq.~(\ref{eq:gw}) are not chirally invariant in 
the canonical sense), but rather be satisfied with a modified form of the 
chiral symmetry. The Nielsen-Ninomyia theorem~\cite{nn} is then circumvented 
altogether, and it is possible to define chiral fermions without doublers or 
breaking of flavor symmetry or other unpleasant drawbacks. The chiral 
transformations (\ref{eq:symmetry}) depend also on the interaction, but this 
does not forbid the non-perturbative construction of chiral gauge theories on 
the lattice with exact gauge invariance~\cite{l3,l1}. For recent reviews on 
the interesting properties of fermions satisfying the Ginsparg-Wilson 
relation, see Refs.~\cite{n,l1,n1}. 

We will be interested in the following in performing chiral-invariant 
computations by using the overlap-Dirac operator (\ref{eq:dn}). This has 
been proven to be local~\footnote{The locality is not meant here to be strict 
locality, but in the larger sense that the strength of the interaction decays 
exponentially with the distance.}~\cite{hjl}, and although simulations with 
the Neuberger operator look computationally very demanding, progress is under
way~\cite{hjl,sim}. There has been activity also on the analytic side, and 
some 1-loop calculations with overlap fermions have been already carried 
out~\cite{ky,ikny,apv,afpv}. The most recent and advanced calculations have
featured the relation between the $\Lambda$ parameter in the lattice scheme 
defined by the overlap operator and in the $\overline{\rm MS}$ 
scheme~\cite{apv}, and the renormalization factors of the quark bilinears 
$\bar{\psi} \Gamma \psi$~\cite{afpv}. 

In this paper we present the calculation in lattice QCD of the renormalization
factors of a few operators which measure the lowest two moments of non-singlet
quark distributions. The generic operators which measure their
$n$-th moments are $\bar{\psi} \gamma_\mu D_{\mu_1} \dots D_{\mu_n} \psi$ 
and $\bar{\psi} \gamma_\mu \gamma_5 D_{\mu_1} \dots D_{\mu_n} \psi$ for 
unpolarized and polarized Structure Functions respectively.
Contrary to what happens with Wilson fermions, in the present case, thanks to 
the exact chiral symmetry that we maintain, the renormalization constants for 
every pair of corresponding unpolarized and polarized operators which differ
by a $\gamma_5$ matrix are the same.

This paper is organized as follows: in Sect.~2 we introduce the various 
operators that we have studied, in Sect.~3 their renormalization on the 
lattice is discussed, and in Sect.~4 some details about the perturbative 
calculations are given. Finally, in Sect.~5 we present our results. In the
Appendices one can find the Feynman rules that we have used as well as some
analytic results.

\section{Moments of Structure Functions}

The hadronic physics in Deep Inelastic Scattering is contained in the 
matrix elements of the T-product of the hadronic currents
\begin{equation}
-i \int d^4x \, e^{iqx} \, 
\langle p | T (J_\mu (x) J_\nu (0)) | p \rangle ,
\end{equation}
$q$ being the momentum transfer and $p$ the target momentum.
An operator product expansion on the light cone of the kind
\begin{equation}
J(x) J(0) \sim \sum_{n,i,l} C^{n,i}_l(x^{2}) \, x^{\mu_1} \cdots x^{\mu_n}
O^{(n,i)}_{\mu_{1} \cdots \mu_{n}}(0) 
\label{eq:ope}
\end{equation}
describes the physics in the Bjorken limit, in which scaling is reached 
and the Structure Functions depend only on the Bjorken variable 
$x_B = -q^2/(2 p \cdot q)$. \linebreak 
The moments of the Structure Functions, which measure the moments of quark
distributions, are directly connected to the forward matrix elements of the 
local operators appearing in the light-cone expansion. Since the divergence 
of the Wilson coefficients is governed by the twist (dimension minus spin) of 
the corresponding operators, infinite towers of operators of increasing 
dimension and spin but with the same twist (dimension minus spin) appear at a 
given order in $1/q^2$. The dominant contribution is given by twist two, 
which in the flavor non-singlet case means the symmetric traceless operators
\begin{eqnarray}
O_{\mu \mu_1 \cdots \mu_n} 
&=& \bar{\psi} \gamma_{\{\mu} D_{\mu_1} \cdots D_{\mu_n\}} \frac{\lambda^a}{2}
\psi \\
O^{(5)}_{\mu \mu_1 \cdots \mu_n} 
&=& \bar{\psi} \gamma_{\{\mu} \gamma_5 D_{\mu_1} \cdots D_{\mu_n\}} 
\frac{\lambda^a}{2}\psi ,
\end{eqnarray}
where $\lambda^a$ are flavor matrices (which in the following will be omitted 
and implicitly understood). These operators measure the moment of unpolarized
and polarized Structure Functions respectively, that is the moments 
$\langle x_B^n \rangle$ of quark momentum distributions and the moments 
$\langle (\Delta x_B)^n \rangle$ of quark helicity distributions, at leading 
twist. Higher-twist operators give the sub-dominant contributions, and in 
particular twist-4 operators (which include also 4-fermion operators) measure 
the $1/q^2$ corrections to these moments~\cite{twist4}. 

Since we are considering flavor non-singlet operators, there can be no mixing 
with operators like 
${\rm Tr} \sum_\rho F_{\mu_1 \rho} D_{\mu_2} \cdots F_{\rho \mu_n}$ and
${\rm Tr} \sum_\rho \widetilde{F}_{\mu_1 \rho} D_{\mu_2} \cdots 
F_{\rho \mu_n}$ which measure the gluon distributions. Mixing with these
operators is prohibited even when one considers unquenched calculations. 
Singlet quark operators instead do mix with gluon operators. For more detailed 
discussions of Deep Inelastic Scattering on the lattice, see 
Refs.~\cite{cr,bbcr,gea,gea2,calcoli}. 

The operators in the light-cone expansion (\ref{eq:ope}) of which we compute 
the renormalization factors are the following:
\begin{eqnarray}
O_{v_2,d} &=& \bar{\psi} \gamma_{\{1} D_{4\}} \psi \label{eq:1ud} \\
O_{v_2,e} &=& \bar{\psi} \gamma_4 D_4 \psi 
               -\frac{1}{3} \sum_{i=1}^3  \bar{\psi} \gamma_i D_i \psi ,
\end{eqnarray}
which measure the first moment of quark momentum distributions,
\begin{eqnarray}
O_{a_2,d} &=& \bar{\psi} \gamma_{\{1} \gamma_5 D_{4\}} \psi \label{eq:1pd} \\
O_{a_2,e} &=& \bar{\psi} \gamma_4 \gamma_5 D_4 \psi 
            -\frac{1}{3} \sum_{i=1}^3  \bar{\psi} \gamma_i \gamma_5 D_i \psi ,
\end{eqnarray}
which measure the first moment of quark helicity distributions,
\begin{eqnarray}
O_{v_3,d} &=& \bar{\psi} \gamma_{\{4} D_1 D_{2\}} \psi \label{eq:2ud} \\
O_{v_3,e} &=& \bar{\psi} \gamma_{\{4} D_1 D_{1\}} \psi 
         -\frac{1}{2} \sum_{i=2}^3  \bar{\psi} \gamma_{\{4} D_i D_{i\}} \psi ,
\label{eq:2ue}
\end{eqnarray}
which measure the second moment of quark momentum distributions, and
\begin{eqnarray}
O_{a_3,d} &=& \bar{\psi} \gamma_{\{4} \gamma_5 D_1 D_{2\}} \psi \\
O_{a_3,e} &=& \bar{\psi} \gamma_{\{4} \gamma_5 D_1 D_{1\}} \psi 
 -\frac{1}{2} \sum_{i=2}^3  \bar{\psi} \gamma_{\{4} \gamma_5 D_i D_{i\}} \psi ,
\label{eq:2pe}
\end{eqnarray}
which measure the second moment of quark helicity distributions.
Symmetrization in all Lorentz indices is understood. We use 
$\stackrel{\phantom{\rightarrow}}{D}=\stackrel{\rightarrow}{D}
-\stackrel{\leftarrow}{D}$, with the following lattice discretizations:
\begin{eqnarray}
\stackrel{\rightarrow}{D_\mu} \psi (x)  &=& \frac{1}{2a} \Big[U (x,\mu) 
\psi(x+a\hat{\mu}) -U^\dagger (x-a\hat{\mu},\mu) \psi(x-a\hat{\mu}) \Big] \\
\bar{\psi} (x) \stackrel{\leftarrow}{D_\mu} &=& \frac{1}{2a} \Big[
\bar{\psi} (x+a\hat{\mu}) U^\dagger (x,\mu) - \bar{\psi}(x-a\hat{\mu}) 
U(x-a\hat{\mu},\mu) \Big] .
\end{eqnarray}

We have chosen the Lorentz indices of each operator appearing in the continuum
expansion in two different ways, so that they fall in two different 
representations of the hypercubic group (the symmetry group of the lattice, 
remnant of the Lorentz symmetry), and on the lattice they will then 
renormalize in a different way. The representations where the indices are 
all different from each other are least likely to mix with other operators, 
however one needs more components of the hadron momentum different from zero 
when simulating the corresponding matrix elements, and this can lead to more 
lattice artifacts. Sometimes one of the representations can be easier to 
handle in practice, depending on the trade-off between the amount of mixing 
and the number of non-zero momentum components. For example, in the case of 
the second moment the operator (\ref{eq:2ud}) is multiplicatively 
renormalized, but it can be more advantageous to use the other representation 
(operator (\ref{eq:2ue})) when it is important that fewer components of the 
momenta are different from zero, although in this case one has then to deal 
with a mixing.

The operator (\ref{eq:2ue}) (together with the corresponding polarized 
(\ref{eq:2pe})) is the only one in the list above that is not multiplicatively
renormalized on the lattice. What happens is that that the two operators
\begin{eqnarray}
O_A &=&\bar{\psi} \gamma_4 D_1 D_1 \psi 
           -\frac{1}{2} \sum_{i=2}^3  \bar{\psi} \gamma_4 D_i D_i \psi , \\
O_B &=&\bar{\psi} \gamma_1 D_4 D_1 \psi +\bar{\psi} \gamma_1 D_1 D_4 \psi
           -\frac{1}{2} \sum_{i=2}^3  \bar{\psi} \gamma_i D_4 D_i \psi 
           -\frac{1}{2} \sum_{i=2}^3  \bar{\psi} \gamma_i D_i D_4 \psi , 
\end{eqnarray}
which are not symmetrized in their indices, do not receive the same 1-loop 
corrections on the lattice~\cite{bbcr}, and so the operator (\ref{eq:2ue}),
\begin{equation}
O_{v_3,e} = \frac{1}{3} \Big( O_A + O_B \Big) ,
\end{equation}
does not go into itself under lattice renormalization, because the symmetric
combination is lost. In the polarized case, an analogous situation occurs for
\begin{eqnarray}
O_{A^5} &=&\bar{\psi} \gamma_4 \gamma_5 D_1 D_1 \psi 
       -\frac{1}{2} \sum_{i=2}^3  \bar{\psi} \gamma_4 \gamma_5 D_i D_i \psi ,\\
O_{B^5} &=&\bar{\psi} \gamma_1 \gamma_5 D_4 D_1 \psi 
      +\bar{\psi} \gamma_1 \gamma_5 D_1 D_4 \psi \\
 &&\quad -\frac{1}{2} \sum_{i=2}^3  \bar{\psi} \gamma_i \gamma_5 D_4 D_i \psi 
         -\frac{1}{2} \sum_{i=2}^3  \bar{\psi} \gamma_i \gamma_5 D_i D_4 \psi .
\nonumber
\end{eqnarray}
A detailed discussion of these effects can be found in Sect.~5.  

We mention that for the second moment there exists also a third independent 
lattice representation, in which all indices are equal. However, it mixes 
with a lower dimensional operator with a power-divergent coefficient 
($\sim 1/a^2$), and so we do not consider it here.

\section{Renormalization}

To relate the numbers obtained from Monte Carlo simulations to physical 
continuum quantities, a lattice renormalization of the relevant matrix 
elements is necessary. The connection is given by
\begin{equation}
\langle O_i^{\rm cont} \rangle = \sum_j \Bigg( \delta_{ij} 
-\frac{g_0^2}{16 \pi^2} \Big( R_{ij}^{\rm lat} -R_{ij}^{\rm cont} \Big) \Bigg) 
\cdot \langle O_j^{\rm lat} \rangle ,
\label{eq:matching}
\end{equation}
where
\begin{equation}
\langle O_i^{\rm cont,lat} \rangle = \sum_j \Bigg( \delta_{ij} 
+ \frac{g_0^2}{16 \pi^2} R_{ij}^{\rm cont,lat} \Bigg) 
\cdot \langle O_j^{\rm tree} \rangle
\end{equation}
are the continuum and lattice 1--loop expressions respectively, and the 
tree-level matrix element is the same in both cases. The differences 
$\Delta R_{ij} = R_{ij}^{\rm lat} - R_{ij}^{\rm cont}$ enter then in the 
renormalization factors
\begin{equation}
Z_{ij} (a\mu,g_0)=\delta_{ij} -\frac{g_0^2}{16 \pi^2} \Delta R_{ij}  (a\mu)
\end{equation}
which connect the lattice to the continuum. These renormalization factors 
are independent of the state, depend only on the scale $a\mu$ and are 
gauge-invariant. Using them, a matrix element obtained from Monte Carlo 
simulations can be renormalized to a continuum scheme.

Although there are in principle also non-perturbative methods with which one 
can determine the renormalization factors, perturbation theory is still 
important. In fact, it can happen that for some operators a window for the 
non-perturbative signal is difficult to obtain, and then perturbation theory 
remains the only possibility of computing the relevant renormalization factors.
For the Neuberger operator, extracting the non-perturbative renormalization 
factors in addition to simulating the bare matrix elements could turn out 
to be computationally very expensive. In general, perturbative lattice 
renormalization is important as a hint and a guide for non--perturbative 
renormalization studies, and even more when mixings are present, which are 
generally more intricate than in the continuum case, and more transparent 
when looked at in perturbation theory, especially if some amounts of mixings 
are small. Perturbative renormalization can in any case be very useful in 
checking and understanding results obtained with non-perturbative methods.

In the continuum we renormalize the operators in the $\overline{\rm{MS}}$
scheme. As perturbative renormalization condition on the lattice we impose 
that the 1-loop amputated matrix elements at a certain reference scale $\mu$ 
are equal to the corresponding bare tree--level quantities. For lattice matrix
elements of multiplicatively renormalized operators computed between 
one-quark states this condition means
\begin{eqnarray}
\langle p |O^{\rm lat}(\mu) | p \rangle \Big|_{p^2=\mu^2} & = & Z_O 
(a\mu, g_0) \cdot Z^{-1}_\psi (a\mu, g_0) \cdot \langle p |O^{(0)}(a) 
| p  \rangle \Big|^{\rm 1-loop}_{p^2=\mu^2} \nonumber \\
& = & \langle p |O^{(0)}(a)   | p  \rangle 
\Big|^{\rm tree}_{p^2=\mu^2},
\end{eqnarray}
where $Z_\psi$ is the wave-function renormalization, computed from the
quark self-energy.

\begin{figure}[btp]
  \begin{center}
    \epsfig{file=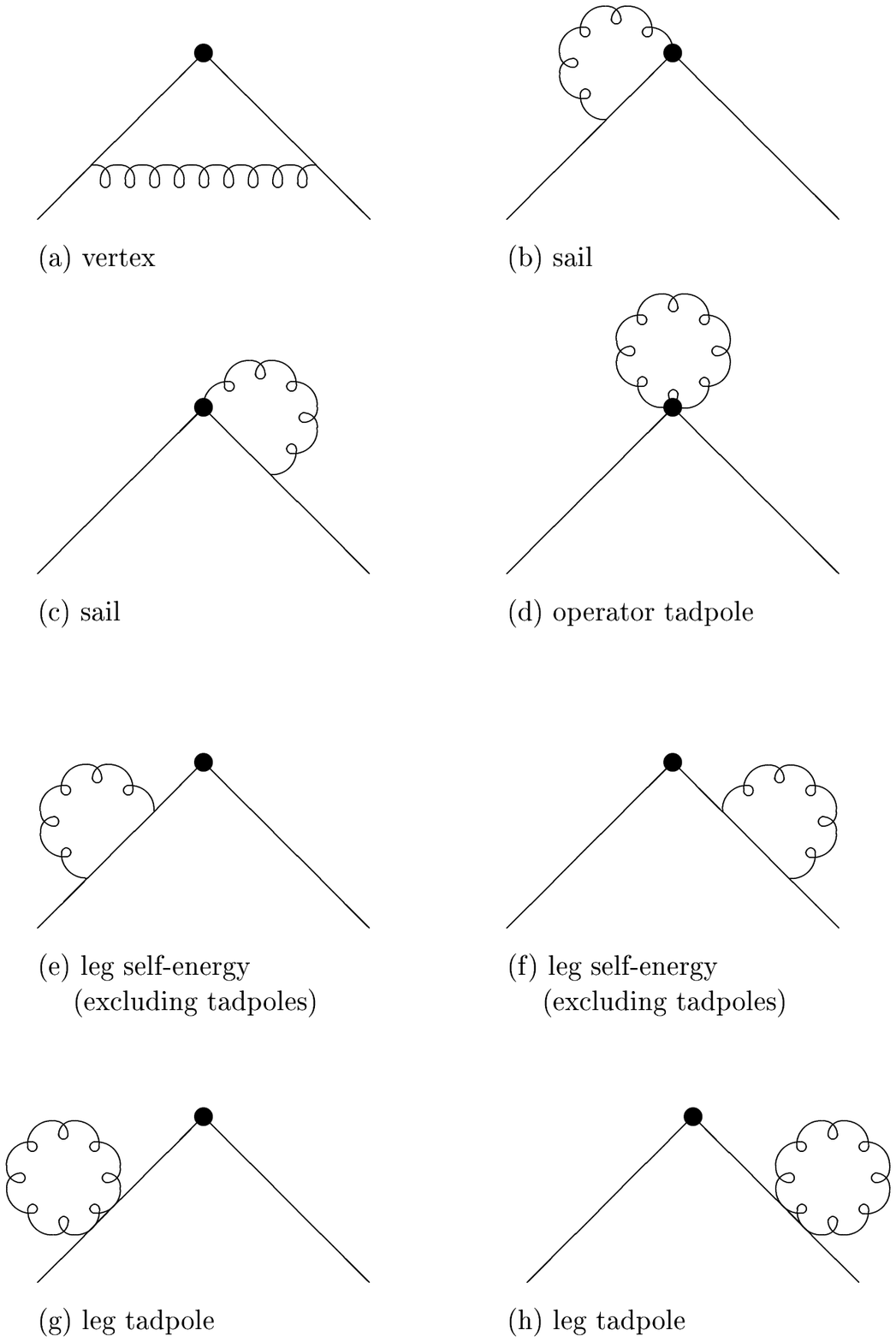, width=.99\linewidth}
    \caption{The graphs contributing to the 1-loop renormalization factors of 
       the matrix elements $\langle p| O |p \rangle$. The operator insertion 
       is indicated by a circle.}
\vspace{0.3cm}
    \label{fig:diagrams}
  \end{center}
\end{figure}

The 1-loop lattice matrix elements of the operators we consider here, which 
are at most logarithmically divergent, will have the form 
\begin{eqnarray}
\langle p |O^{(0)}(a) | p  \rangle \Big|^{\rm 1-loop} &=&
\langle p |O^{(0)}(a)   | p  \rangle \Big|^{\rm tree} \times \\
&& \qquad \Bigg( 1 + \frac{g_0^2}{16\pi^2} C_F \Big( \gamma_O \log a^2 p^2 
+V_O + T_O +2S \Big) \Bigg) \nonumber ,
\end{eqnarray}
where $V_O$ is the finite contribution of the vertex and sails diagrams 
(a, b and c in Fig.~1), $T_O$ refers to the tadpole arising from the operator 
(d in Fig.~1), S is the finite contribution (proportional to $i p \!\!\! /$)
of the quark self-energy of one leg, including the leg tadpole (e and g, 
or f and h, in Fig.~1), and $C_F=\frac{N_c^2-1}{2N_c}$. 
The $Z_O$ factor for an operator $O$ will then be given by
\begin{equation}
Z_O (a\mu, g_0) = 
1 - \frac{g_0^2}{16\pi^2} C_F \Bigg( \gamma_O \log a^2 \mu^2 +B_O \Bigg) ,
\end{equation}
with
\begin{equation}
B_O = V_O + T_O + S.
\end{equation}
We will call ``proper'' contributions the ones that exclude the self-energy
diagrams. They correspond to the diagrams a-d in Fig.~1.

Although the theory defined by the overlap-Dirac operator (\ref{eq:dn}) has 
an exact chiral symmetry and no lattice artifacts of order $a$ are present 
in the action, matrix elements of operators still possess corrections of order 
$a$ and therefore they need to be improved. The improved operator corresponding
to $\bar{\psi} \gamma_{\{\mu} D_{\mu_1} \cdots D_{\mu_n\}} \psi$
is~\cite{cghrs}
\begin{equation}
\bar{\psi} \Big(1-\frac{1}{2} a D_N\Big) \, \gamma_{\{\mu} D_{\mu_1} 
\cdots D_{\mu_n\}} \, \Big(1-\frac{1}{2} a D_N\Big) \psi,
\label{eq:opimp}
\end{equation}
and the big bonus with overlap fermions is that the renormalization factors 
for the improved operator and for the unimproved operator without the 
rotations $ (1-\frac{1}{2} a D_N)$ are the same. What happens is that in 
1-loop amplitudes a factor $D_N$ can combine with a quark propagator, but 
since it has an $a$ in front and (contrary to the Wilson case) there is no 
$1/a$ factor in the propagator, as additive mass renormalization is forbidden 
by chiral symmetry, the contribution of $D_N$ to the renormalization factors 
is zero~\cite{afpv}. Thus, we can simulate the improved operator 
(\ref{eq:opimp}) and renormalize it as if it were the unimproved operator, 
which saves a lot in the perturbative calculations. 

Chiral symmetry, in addition to avoiding the extrapolations to zero quark
mass in the simulations, gives also useful constraints on the mixing patterns 
and on the values of the renormalization factors. In general, mixings that in 
the Wilson case were allowed by the breaking of chirality are now forbidden, 
although this has less relevance here than in other problems like in weak 
interactions~\cite{cg}. An important consequence here is that, similarly to 
what happens in the case of the bilinears where we have $Z_S=Z_P$ and 
$Z_V=Z_A$~\cite{afpv}, chiral symmetry forces pairs of corresponding 
unpolarized and polarized operators, like $O_{v_2,d}$ and $O_{a_2,d}$, 
to have the same renormalization constants~\footnote{This is strictly true for
improved operators like in Eq.~(\ref{eq:opimp}), which transform like chiral
multiplets, but since they have the same renormalization constants as the 
unimproved operators, these symmetry relations are preserved.}. 

\section{The computations}

The interaction vertices and the propagator of the overlap-Dirac operator 
are much more complicated than the ones in the Wilson formulation, and this 
causes the computations to be rather cumbersome (see Appendices). Computer
programs need then to be introduced, and we have performed the calculations 
using an ensemble of routines written in the symbolic manipulation language 
FORM. These routines are an extension of the ones used to do calculations
with the Wilson action in various occasions~\cite{calcoli,calcoli2}.

The outputs of the FORM codes, which correspond to the results of the analytic 
calculations, are fed to Fortran programs which perform the numerical
integrations. To treat the divergent integrals we use the Kawai 
method~\cite{kns}, in which they are split in integrals independent of 
external momenta plus continuum integrals. Divergent lattice integrals which 
contain both gluon and overlap-quark propagators are reduced to divergent
integrals containing only gluon propagators, which are computed in a exact 
way as in~\cite{cmp}, plus a finite rest.

Many checks have been performed on our results. We did the computations 
in a general covariant gauge~\footnote{The gluon propagator 
that we use is 
\begin{equation}
G_{\mu\nu}(k) = \frac{1}{4\sum_\rho \sin^2 \frac{k_\rho}{2}}
\Bigg( \delta_{\mu\nu} - (1-\alpha) \frac{4 \sin \frac{k_\mu}{2} 
\sin \frac{k_\nu}{2}}{4\sum_\lambda \sin^2 \frac{k_\lambda}{2}} \Bigg) .
\end{equation}
}, and the cancellations of the gauge-dependent parts between the continuum 
scheme and the lattice results is a strong check on the good behavior of the 
FORM codes, as well as of the integration routines.

We have used the codes to compute also the renormalization constants for all 
quark bilinears, and we agree with the results given in Feynman-gauge by 
Alexandrou et al.~\cite{afpv}, as well as with their self-energy. In the case 
of the leg tadpole and of the vertex graph of the scalar current 
$\bar{\psi} \psi$ we have also made the computations entirely by hand 
(although only in the Feynman gauge), and successfully checked them against 
the FORM output expressions. These analytic results are reported in Appendix 
B. We tried to do a calculation by hand for a first moment operator and for 
the rest of the leg self-energy, but they involved a huge amount of 
manipulations. We rely also on the vast amount of results which have been 
produced in different occasions using these FORM codes limited to the Wilson 
formulation~\cite{calcoli,calcoli2}, and these results have by now a certain 
number of cross-checks. With overlap fermions we need to introduce a new 
propagator and new vertices, but the operators are unaltered, and the same
Wilson expressions for their expansion in $a$ can be used here. Many of the 
routines (for example the gamma algebra reduction) are also exactly the same 
as in the Wilson case.

We checked in all cases that the polarized operators have the same $Z$s as 
the corresponding unpolarized operators. This is a rather strong check, as 
the analytic results are widely different for the two cases, and it is only 
after performing the numerical integrations and seeing that the difference 
of the two results is much smaller than the precision of our integrals that 
we able to say that the renormalization factors are indeed equal. 
Finally, since our codes are able to do calculations both in Dimensional 
Regularization and using a mass regularization, we also checked that using 
two different intermediate regularizations while treating the divergences 
with the Kawai method~\cite{kns} we get the same results.

We obtain the $B_O$ constants with 5 digits after the decimal points, by 
doing the numerical integrations on a $40^4$ regular grid and using the 
method proposed by L\"uscher and Weisz in~\cite{l4} to accelerate 
convergence. We have checked that these digits do not vary when a $60^4$ 
integration grid is used. In any case, to get more significant digits, grids 
of $80^4$ or $100^4$ points would still not be enough, and they would require 
an enormous computational effort, as a typical FORM output for the 
second moment operators already contains thousands of terms.
Comparing our results for the self-energy and the bilinears with the numbers 
in Ref.~\cite{afpv}, we can see that a few times a difference of one unit on 
the fifth digit can be noticed, and this agrees with our estimate of errors. 
In that paper the results are given in terms of the constants 
$b=\frac{B}{16\pi^2}$, and this explains the presence of two more significant 
digits after the decimal point.

\section{Results}

We give in this Section the renormalization factors of the operators 
considered in this work. That these factors are identical for pairs of 
corresponding unpolarized and polarized operators has been explicitly verified
for all operators. We give then the numerical results only for the unpolarized
operators, for $r=1$. 

We first consider the 1-loop contributions of the proper diagrams (a-d in 
Fig.~1), which for the operators that do not mix are
\begin{eqnarray}
O_{v_2,d}^{\rm proper} =\frac{g_0^2}{16\pi^2} C_F && \Bigg[ \Big( \frac{5}{3} 
+ (1-\alpha) \Big) \log a^2 p^2  \nonumber \\ 
&& \qquad + V^{\ \alpha=1}_{v_2,d} - (1-\alpha)\, 6.850272 
+ T_{v_2,d} \Bigg] O_{v_2,d}^{\rm tree} , \nonumber \\
O_{v_2,e}^{\rm proper} =\frac{g_0^2}{16\pi^2} C_F && \Bigg[ \Big( \frac{5}{3} 
+ (1-\alpha) \Big) \log a^2 p^2 \\ 
&& \qquad + V^{\ \alpha=1}_{v_2,e} - (1-\alpha)\, 6.850272
+ T_{v_2,e} \Bigg] O_{v_2,e}^{\rm tree} , \nonumber \\
O_{v_3,d}^{\rm proper} =\frac{g_0^2}{16\pi^2} C_F && \Bigg[ \Big( \frac{19}{6} 
+ (1-\alpha) \Big) \log a^2 p^2  \nonumber \\
&& \qquad +V^{\ \alpha=1}_{v_3,d} - (1-\alpha)\, 7.369693
+ T_{v_3,d} \Bigg] O_{v_3,d}^{\rm tree} \nonumber .
\end{eqnarray}
The contributions of the sails and vertices, $V_O$, have been for convenience
separated in the Feynman gauge results ($V^{\ \alpha=1}_{v_2,d}$, 
$V^{\ \alpha=1}_{v_2,e}$ and $V^{\ \alpha=1}_{v_3,d}$), tabulated in Table 1 
for various values of the parameter $\rho$, and the remaining parts 
proportional to $(1-\alpha)$, which are instead independent of $\rho$. 
The analytic expressions of the latter are very complicated functions of 
$\rho$ containing thousands of terms, and the numerical cancellation of this 
dependence is a rather non-trivial check of our computations. These numbers 
are also independent of the lattice representation, and furthermore they have 
the same value as with the Wilson action~\cite{calcoli2}, so they seem to be 
to a certain extent independent of the particular fermion action chosen.

\begin{table}[hbtp]
\begin{center}
\vspace{0.5cm}
  \begin{tabular}[btp]{|c||r|r|r|} \hline
$\rho$ & $V_{v_2,d}^{\ \alpha=1}=V_{a_2,d}^{\ \alpha=1}$ &
$V_{v_2,e}^{\ \alpha=1}=V_{a_2,e}^{\ \alpha=1}$ & 
$V_{v_3,d}^{\ \alpha=1}=V_{a_3,d}^{\ \alpha=1}$ \\ \hline
0.2  &  -3.77889  &  -3.50634  &  -7.57984  \\
0.3  &  -3.71038  &  -3.41915  &  -7.45926  \\
0.4  &  -3.65209  &  -3.34178  &  -7.35607  \\
0.5  &  -3.60037  &  -3.27058  &  -7.26432  \\
0.6  &  -3.55327  &  -3.20357  &  -7.18078  \\
0.7  &  -3.50959  &  -3.13952  &  -7.10341  \\
0.8  &  -3.46851  &  -3.07762  &  -7.03088  \\
0.9  &  -3.42946  &  -3.01727  &  -6.96221  \\
1.0  &  -3.39203  &  -2.95803  &  -6.89668  \\
1.1  &  -3.35589  &  -2.89955  &  -6.83375  \\
1.2  &  -3.32077  &  -2.84155  &  -6.77296  \\
1.3  &  -3.28647  &  -2.78380  &  -6.71398  \\
1.4  &  -3.25281  &  -2.72610  &  -6.65650  \\
1.5  &  -3.21964  &  -2.66827  &  -6.60029  \\
1.6  &  -3.18685  &  -2.61015  &  -6.54513  \\
1.7  &  -3.15430  &  -2.55162  &  -6.49084  \\
1.8  &  -3.12190  &  -2.49255  &  -6.43729  \\
\hline
  \end{tabular} 
\vspace{0.5cm}
  \caption{The Feynman-gauge constants $V^{\ \alpha=1}_O$ for the
multiplicatively renormalized operators.}
\end{center}
\vspace{0.5cm}
\end{table}

\begin{table}[hbtp]
\begin{center}
\vspace{0.5cm}
  \begin{tabular}[btp]{|c||l|} \hline
 operator & operator tadpole \\ \hline
$O_{v_2,d},O_{a_2,d}$ & $ T_{v_2,d} = T_{a_2,d} = 
-16\pi^2\frac{Z_0}{2}\Big( 1-\frac{1}{4} (1-\alpha) \Big) $ \\
$O_{v_2,e},O_{a_2,e}$ & $ T_{v_2,e} = T_{a_2,e} = 
-16\pi^2\frac{Z_0}{2}\Big( 1-\frac{1}{4} (1-\alpha) \Big) $ \\
$O_{v_3,d},O_{a_3,d}$ & $ T_{v_3,d} = T_{a_3,d} = 
16 \pi^2 \Big (-Z_0 + (1-\alpha) \, \frac{Z_0}{6}\Big) $ \\ \hline
            & $ T_{AA} = T_{A^5A^5} =  
16 \pi^2 \Big( -2 Z_0 +\frac{1}{8} + (1-\alpha) \Big( 
Z_0 - \frac{1}{8} \Big) \Big) $ \\
$O_{v_3,e},O_{a_3,e}$ & $ T_{AB} = T_{A^5B^5} = 0 $ \\
            & $ T_{BA} =  T_{B^5A^5} = 0 $ \\
            & $  T_{BB} =   T_{B^5B^5} = 
16 \pi^2 \Big (-Z_0 + (1-\alpha) \, \frac{Z_0}{6}\Big) $ \\
\hline
  \end{tabular} 
\vspace{0.5cm}
  \caption{The operator tadpoles for the various operators, where 
$Z_0=0.154933390231$.}
\end{center}
\vspace{0.5cm}
\end{table}

\begin{table}[hbtp]
\begin{center}
\vspace{0.5cm}
  \begin{tabular}[btp]{|c||r|r|r|r|} \hline
  $\rho$ & $V_{AA}^{\ \alpha=1} = V_{A^5A^5}^{\ \alpha=1}$ 
& $V_{BA}^{\ \alpha=1} = V_{B^5A^5}^{\ \alpha=1}$ 
& $V_{AB}^{\ \alpha=1} = V_{A^5B^5}^{\ \alpha=1}$
& $V_{BB}^{\ \alpha=1} = V_{B^5B^5}^{\ \alpha=1}$ \\ \hline
0.2  &  22.83163  &  -36.93402  &  -13.77026  &  10.81988  \\ 
0.3  &  13.03570  &  -24.07048  &   -8.80836  &   4.50553  \\
0.4  &   8.39467  &  -17.86369  &   -6.43263  &   1.50207  \\
0.5  &   5.76489  &  -14.27357  &   -5.06826  &  -0.20460  \\
0.6  &   4.11604  &  -11.96971  &   -4.19847  &  -1.27644  \\
0.7  &   3.01383  &  -10.38860  &   -3.60511  &  -1.99319  \\
0.8  &   2.24472  &   -9.25173  &   -3.18071  &  -2.49274  \\
0.9  &   1.69205  &   -8.40616  &   -2.86648  &  -2.85062  \\
1.0  &   1.28701  &   -7.76116  &   -2.62765  &  -3.11148  \\
1.1  &   0.98657  &   -7.25963  &   -2.44241  &  -3.30329  \\
1.2  &   0.76254  &   -6.86397  &   -2.29645  &  -3.44445  \\
1.3  &   0.59577  &   -6.54843  &   -2.18002  &  -3.54747  \\
1.4  &   0.47280  &   -6.29483  &   -2.08624  &  -3.62114  \\
1.5  &   0.38399  &   -6.08998  &   -2.01015  &  -3.67183  \\
1.6  &   0.32223  &   -5.92413  &   -1.94810  &  -3.70420  \\ 
1.7  &   0.28218  &   -5.78989  &   -1.89733  &  -3.72177  \\
1.8  &   0.25980  &   -5.68156  &   -1.85572  &  -3.72723  \\
\hline
  \end{tabular} 
\vspace{0.5cm}
  \caption{The Feynman-gauge constants $V^{\ \alpha=1}_O$ for the operators
$O_A$ and $O_B$ related to $O_{v_3,e}$.}
\end{center}
\vspace{0.5cm}
\end{table}

\begin{table}[hbtp]
\begin{center}
\vspace{0.5cm}
  \begin{tabular}[btp]{|r|r||r|r||r|r|} \hline
$\rho$ & S &  $\rho$ & S &  $\rho$ & S  \\ 
\hline
0.2 & -240.59963 &  0.8 & -49.99380 &  1.4 & -23.76598  \\
0.3 & -155.41069 &  0.9 & -43.10649 &  1.5 & -21.50111  \\
0.4 & -112.98999 &  1.0 & -37.63063 &  1.6 & -19.53531  \\
0.5 &  -87.65282 &  1.1 & -33.17935 &  1.7 & -17.81537  \\
0.6 &  -70.84428 &  1.2 & -29.49505 &  1.8 & -16.29999  \\
0.7 &  -58.90122 &  1.3 & -26.39961 &  1.9 & -14.95658  \\
\hline
  \end{tabular} 
\vspace{0.5cm}
\caption{1-loop results for the quark self-energy (including the leg tadpole).}
\end{center}
\vspace{0.5cm}
\end{table}

The contributions $T_O$ of the operator tadpoles (d in Fig.~1) are shown 
in Table 2, which contains also their results for the other operator, 
$O_{v_3,e}$. However, since this operator is not multiplicatively 
renormalized on the lattice, we consider in its place the two operators 
$O_A$ and $O_B$ introduced in Sect.~2. They mix with each other and we can 
write their renormalization as~\footnote{In Refs.~\cite{gea} another choice 
for the two operators that mix was made, and $O_{v_3,e}$ and an operator of 
mixed symmetry were considered instead of $O_A$ and $O_B$.}
\begin{eqnarray}
\widehat{O}_A &=& Z_{AA} O_A + Z_{AB} O_B \\
\widehat{O}_B &=& Z_{BA} O_A + Z_{BB} O_B \nonumber .
\end{eqnarray}
In terms of the bare operators $O_A$ and $O_B$, the renormalized operator 
$\widehat{O}_{v_3,e}$ appearing in the DIS light-cone expansion will be
\begin{equation}
\widehat{O}_{v_3,e} = \frac{1}{3} \Big( \widehat{O}_A + \widehat{O}_B \Big)
= \frac{1}{3} (Z_{AA} + Z_{BA}) \, O_A + \frac{1}{3} (Z_{AB} + Z_{BB}) \, O_B ,
\end{equation}
and since on the lattice $Z_{AA}+Z_{BA} \neq Z_{AB}+Z_{BB}$, the result is 
that $O_{v_3,e}$ is not multiplicatively renormalized. In fact, the 1-loop 
contributions coming from the proper diagrams are in this case
\begin{eqnarray}
O_A^{\rm proper} = \frac{g_0^2}{16\pi^2} C_F && \Bigg[ \Big( \frac{7}{6} 
+ (1-\alpha) \Big) \log a^2 p^2 \nonumber \\ 
&& \qquad + V_{AA}^{\ \alpha=1} - (1-\alpha)\, 8.685568
+ T_{AA} \Bigg] O_A^{\rm tree} \nonumber \\
+ \frac{g_0^2}{16\pi^2} C_F && \Bigg[ \log a^2 p^2 + V_{AB}^{\ \alpha=1} 
+ \frac{1}{3} (1-\alpha) + T_{AB} \Bigg] O_B^{\rm tree}  \nonumber \\ 
O_B^{\rm proper} =  \frac{g_0^2}{16\pi^2} C_F && \Bigg[ 2\log a^2 p^2 
+ V_{BA}^{\ \alpha=1} +\frac{2}{3} (1-\alpha) + T_{BA} \Bigg] O_A^{\rm tree} \\
+ \frac{g_0^2}{16\pi^2} C_F && \Bigg[ \Big( \frac{13}{6} 
+ (1-\alpha) \Big) \log a^2 p^2 \nonumber \\ 
&& \qquad + V_{BB}^{\ \alpha=1} - (1-\alpha) \, 7.703026
+ T_{BB} \Bigg] O_B^{\rm tree} \nonumber,
\end{eqnarray}
where the finite contributions of vertex and sails in the Feynman gauge are 
shown in Table 3. 

We stress again that this mixing is a pure lattice effect, deriving from the
breaking of the Lorentz group to the hypercubic group. In the continuum we do 
have $Z_{AA}+Z_{BA} = Z_{AB}+Z_{BB}$, and thus the operator $O_{v_3,e}$ is 
multiplicatively renormalized, and has the same $Z$ as the operator with 
different indices $O_{v_3,d}$, as they both belong to the same representation 
of the Lorentz group. We remind also that in the polarized case one encounters
exactly the same situation with the same numerical results, i.e. 
$Z_{AA} = Z_{A^5A^5}$, $Z_{AB} = Z_{A^5B^5}$ etc., as we have explicitly 
verified. 

To complete the computation of the renormalization factors, we have now to 
add to the proper diagrams the 1-loop amplitudes of the self-energy and 
tadpole of one leg which are proportional to $i p \!\!\! /$,
\begin{equation}
\Sigma_1 = \frac{g_0^2}{16\pi^2} C_F \Bigg[ \alpha \log a^2 p^2  
+S^{\ \alpha=1} +(1-\alpha) \, 4.792010 \Bigg],
\end{equation}
where the Feynman-gauge finite results $S^{\ \alpha=1}$ are given in Table 4. 
Putting everything together, we get the expressions of the renormalized 
operators on the lattice for overlap fermions, which for $\rho=1$ are:
\begin{eqnarray}
\widehat{O}_{v_2,d} &=&\Bigg[ 1 -\frac{g_0^2}{16\pi^2} C_F 
\Bigg(\frac{8}{3} \log a^2 \mu^2 
-53.25571 +(1-\alpha) \Bigg) \Bigg] O_{v_2,d}^{\rm tree} \nonumber \\ 
\widehat{O}_{v_2,e} &=&\Bigg[ 1 -\frac{g_0^2}{16\pi^2} C_F 
\Bigg(\frac{8}{3} \log a^2 \mu^2 
-52.82171  +(1-\alpha) \Bigg) \Bigg] O_{v_2,e}^{\rm tree} \nonumber \\ 
\widehat{O}_{v_3,d} &=& \Bigg[ 1 -\frac{g_0^2}{16\pi^2} C_F 
\Bigg(\frac{25}{6} \log a^2 \mu^2 
-68.99341  +\frac{3}{2} \, (1-\alpha) \Bigg) \Bigg] O_{v_3,d}^{\rm tree} 
\nonumber \\ 
\widehat{O}_{v_3,e} &=& \frac{1}{3} \Bigg[ 1 -\frac{g_0^2}{16\pi^2} C_F 
\Bigg(\frac{25}{6} \log a^2 \mu^2 
-73.29777  +\frac{3}{2} \, (1-\alpha) \Bigg) \Bigg] O_A^{\rm tree} \\ 
                    && + \frac{1}{3} \Bigg[ 1 -\frac{g_0^2}{16\pi^2} C_F 
\Bigg(\frac{25}{6} \log a^2 \mu^2 
-67.83586  +\frac{3}{2} \, (1-\alpha) \Bigg) \Bigg] O_B^{\rm tree} . \nonumber 
\end{eqnarray}

We give here for comparison the numerical values obtained with the usual 
Wilson action, without any improvement. We have computed them again and 
checked them with the results in the 
literature~\cite{cr,bbcr,gea,gea2,calcoli,calcoli2}. 
In this case however chiral invariance is broken and the renormalization 
factors for the unpolarized operators 
\begin{eqnarray}
\widehat{O}^{\rm Wilson}_{v_2,d} 
&=&\Bigg[ 1 -\frac{g_0^2}{16\pi^2} C_F \Bigg(\frac{8}{3} \log a^2 \mu^2 
-3.16486  +(1-\alpha) \Bigg) \Bigg] O_{v_2,d}^{\rm tree} \nonumber \\ 
\widehat{O}^{\rm Wilson}_{v_2,e} 
&=&\Bigg[ 1 -\frac{g_0^2}{16\pi^2} C_F \Bigg(\frac{8}{3} \log a^2 \mu^2 
-1.88259  +(1-\alpha) \Bigg) \Bigg] O_{v_2,e}^{\rm tree} \nonumber \\ 
\widehat{O}^{\rm Wilson}_{v_3,d} &=& \Bigg[ 1 -\frac{g_0^2}{16\pi^2} C_F
\Bigg( \frac{25}{6} \log a^2 \mu^2 
-19.00763  +\frac{3}{2} \, (1-\alpha) \Bigg) \Bigg] O_{v_3,d}^{\rm tree}
\nonumber \\ 
\widehat{O}^{\rm Wilson}_{v_3,e} &=& \frac{1}{3} 
\Bigg[ 1 -\frac{g_0^2}{16\pi^2} C_F \Bigg( \frac{25}{6} \log a^2 \mu^2 
-21.78271  +\frac{3}{2} \, (1-\alpha) \Bigg) \Bigg] O_A^{\rm tree} \\ 
                    &&+ \frac{1}{3}  
\Bigg[ 1 - \frac{g_0^2}{16\pi^2} C_F \Bigg( \frac{25}{6} \log a^2 \mu^2 
-18.46640  +\frac{3}{2} \, (1-\alpha) \Bigg) \Bigg] O_B^{\rm tree}  \nonumber
\end{eqnarray}
differ from the ones of the polarized operators 
\begin{eqnarray}
\widehat{O}^{\rm Wilson}_{a_2,d} 
&=&\Bigg[ 1 -\frac{g_0^2}{16\pi^2} C_F \Bigg(\frac{8}{3} \log a^2 \mu^2 
-4.09933  +(1-\alpha) \Bigg) \Bigg] O_{a_2,d}^{\rm tree} \nonumber \\ 
\widehat{O}^{\rm Wilson}_{a_2,e} 
&=&\Bigg[ 1 -\frac{g_0^2}{16\pi^2} C_F \Bigg(\frac{8}{3} \log a^2 \mu^2 
-4.27705  +(1-\alpha) \Bigg) \Bigg] O_{a_2,e}^{\rm tree} \nonumber \\ 
\widehat{O}^{\rm Wilson}_{a_3,d} &=& \Bigg[ 1 -\frac{g_0^2}{16\pi^2} C_F 
\Bigg( \frac{25}{6} \log a^2 \mu^2 
-19.56159  +\frac{3}{2} \, (1-\alpha) \Bigg) \Bigg] O_{a_3,d}^{\rm tree} 
\nonumber \\ 
\widehat{O}^{\rm Wilson}_{a_3,e} &=& \frac{1}{3} 
\Bigg[ 1 -\frac{g_0^2}{16\pi^2} C_F \Bigg( \frac{25}{6} \log a^2 \mu^2 
-22.39940  +\frac{3}{2} \, (1-\alpha) \Bigg) \Bigg] O_{A^5}^{\rm tree} \\ 
                    &&+\frac{1}{3} \Bigg[ 1 - \frac{g_0^2}{16\pi^2} C_F 
\Bigg( \frac{25}{6} \log a^2 \mu^2 
-19.25837  +\frac{3}{2} \, (1-\alpha) \Bigg)\Bigg] O_{B^5}^{\rm tree}.
\nonumber 
\end{eqnarray}
What one can notice is that the values of the renormalization factors for 
overlap fermions are in general much larger than for Wilson fermions. This 
seems to be true for most values of $\rho$, and is to trace largely to the 
quark self-energy and to the operators tadpoles. It can be observed that when 
in the Wilson formulation one improves the theory by adding the clover term 
to the action (with $c_{sw}=1$) and by canceling $O(a)$ effects on the 
operators at tree level, the renormalization factors become also somewhat 
larger, as in the examples below~\cite{cr,bbcr}:
\begin{eqnarray}
\widehat{O}^{\rm Wilson,~imp}_{v_2,d} 
&=&\Bigg[ 1 -\frac{g_0^2}{16\pi^2} C_F \Bigg(\frac{8}{3} \log a^2 \mu^2 
-15.816  +(1-\alpha) \Bigg) \Bigg] O_{v_2,d}^{\rm tree} \nonumber \\ 
\widehat{O}^{\rm Wilson,~imp}_{v_3,d} &=& \Bigg[ 1 -\frac{g_0^2}{16\pi^2} C_F 
\Bigg( \frac{25}{6} \log a^2 \mu^2 
-29.815  +\frac{3}{2} \, (1-\alpha) \Bigg) \Bigg] O_{v_3,d}^{\rm tree} 
\nonumber \\ 
\widehat{O}^{\rm Wilson,~imp}_{v_3,e} &=& \frac{1}{3} 
\Bigg[ 1 -\frac{g_0^2}{16\pi^2} C_F \Bigg( \frac{25}{6} \log a^2 \mu^2 
-39.192  +\frac{3}{2} \, (1-\alpha) \Bigg) \Bigg] O_A^{\rm tree} \\ 
                    && + \frac{1}{3} 
\Bigg[ 1 - \frac{g_0^2}{16\pi^2} C_F \Bigg( \frac{25}{6} \log a^2 \mu^2 
-22.141  +\frac{3}{2} \, (1-\alpha) \Bigg) \Bigg] O_B^{\rm tree}  \nonumber .
\end{eqnarray}
One could then speculate whether the large renormalization factors that we 
have obtained for overlap fermions are related to the fact that the Neuberger
action is improved and the $Z$s correspond to the ones of improved operators. 
A calculation with Wilson fermions with which to compare in this sense our
overlap results would be one in which the operators are improved beyond tree 
level, but although we know in some cases the contributions of the operator
counterterms that are needed for the full improvement, unfortunately not
all their coefficients are yet determined~\cite{calcoli2}.

We give finally the 1-loop results for the various matrix elements in the 
$\overline{\rm{MS}}$ scheme, so that one can complete the connection with 
the continuum as in Eq.~(\ref{eq:matching}). In the continuum there is only 
one renormalization constant for all possible operators (unpolarized and 
polarized) measuring the first moment, and only one for all possible operators
measuring the second moment:
\begin{eqnarray}
\widehat{O}^{\overline{\rm{MS}}}_{v_2} 
&=& \Bigg[ 1 - \frac{g_0^2}{16\pi^2} C_F \Bigg( \frac{8}{3} 
\log \frac{p^2}{\mu^2} -\frac{40}{9} +(1-\alpha) \Bigg) \Bigg] 
O_{v_2}^{\rm tree} \label{eq:msbar}\\ 
\widehat{O}^{\overline{\rm{MS}}}_{v_3} 
&=& \Bigg[ 1 - \frac{g_0^2}{16\pi^2} C_F \Bigg( \frac{25}{6} 
\log \frac{p^2}{\mu^2} -\frac{67}{9} +\frac{3}{2} \, (1-\alpha) \Bigg) \Bigg] 
O_{v_3}^{\rm tree} . \nonumber 
\end{eqnarray}
The connection of overlap lattice fermions with the continuum 
$\overline{\rm{MS}}$ is then given by the gauge-invariant factors
\begin{eqnarray}
\widehat{O}^{\overline{\rm{MS}}}_{v_2,d} 
&=&\Bigg[ 1 -\frac{g_0^2}{16\pi^2} C_F \Bigg(\frac{8}{3} \log a^2 \mu^2 
-48.81127 \Bigg) \Bigg] O^{\rm lat}_{v_2,d} \nonumber \\ 
\widehat{O}^{\overline{\rm{MS}}}_{v_2,e} 
&=&\Bigg[ 1 -\frac{g_0^2}{16\pi^2} C_F \Bigg(\frac{8}{3} \log a^2 \mu^2 
-48.37727 \Bigg) \Bigg] O^{\rm lat}_{v_2,e} \nonumber \\ 
\widehat{O}^{\overline{\rm{MS}}}_{v_3,d} &=& \Bigg[ 1 - \frac{g_0^2}{16\pi^2}
C_F \Bigg(\frac{25}{6} \log a^2 \mu^2 
-61.54897 \Bigg) \Bigg] O^{\rm lat}_{v_3,d} \nonumber \\ 
\widehat{O}^{\overline{\rm{MS}}}_{v_3,e} &=& \frac{1}{3} 
\Bigg[ 1 - \frac{g_0^2}{16\pi^2} C_F \Bigg(\frac{25}{6} \log a^2 \mu^2 
-65.85333 \Bigg) \Bigg] O^{\rm lat}_A \\ 
                    &&+ \frac{1}{3} 
\Bigg[ 1 - \frac{g_0^2}{16\pi^2} C_F \Bigg( \frac{25}{6} \log a^2 \mu^2 
-60.39142 \Bigg) \Bigg] O^{\rm lat}_B . \nonumber 
\end{eqnarray}

It looks increasingly difficult to go to higher moments, as the number of 
terms that are present in the FORM outputs and that need to be numerically 
integrated becomes very large. This has at the moment limited our calculations
to second moment operators, but we hope to be able to compute the 
renormalization of third-moment operators in the near future.

Finally, we remark that all the renormalization constants presented here can 
be considered as computed in the unquenched case, because we limit ourselves 
to 1-loop computations with non-singlet quark operators, where internal
quark loops never have the chance to come to play.

\section*{Acknowledgment}
I have enjoyed stimulating discussions with Robert Edwards and Leonardo 
Giusti. I would also like to thank Mark Alford for critically reading the 
manuscript. Both the FORM and Fortran computations have been done at MIT 
on a few Pentium III PCs running on Linux. This work has been supported 
in part by the U.S. Department of Energy (DOE) under cooperative research 
agreement DE-FC02-94ER40818.

\appendix

\section{Feynman rules for the Neuberger operator}

We give here the Feynman rules which are needed to perform 1-loop calculations 
using the overlap-Dirac operator. An explicit derivations of these rules is 
given in~\cite{ky,ikny}. 

The quark propagator is
\begin{equation}
S(k) = \frac{-i \sum_\mu \gamma_\mu \sin ak_\mu}{2 \rho
\left( \omega(k)+b(k) \right) } + \frac{a}{2 \rho} ,
\end{equation}
where
\begin{eqnarray}
\omega(k) &=& \frac{1}{a} \sqrt{ \sum_\mu \sin^2 ak_\mu + \Bigg( 2r \sum_\mu
\sin^2 \frac{ak_\mu}{2} -\rho \Bigg)^2 } \nonumber \\
b(k) &=& \frac{1}{a} \Bigg( 2r \sum_\mu
\sin^2 \frac{ak_\mu}{2} -\rho \Bigg).
\end{eqnarray}

The vertices needed for 1-loop calculations can be entirely given in terms of 
the vertices of the Wilson action
\begin{eqnarray}
W_{1\mu}(p_1,p_2) &=& - g_0 \Bigg( i \gamma_\mu \cos \frac{a(p_1+p_2)_\mu}{2}
              + r \sin \frac{a(p_1+p_2)_\mu}{2} \Bigg) \\
W_{2\mu}(p_1,p_2) &=& 
- \frac{1}{2} a g_0^2 \Bigg( -i \gamma_\mu \sin \frac{a(p_1+p_2)_\mu}{2}
              + r \cos \frac{a(p_1+p_2)_\mu}{2} \Bigg) 
\end{eqnarray}
(where $p_1$ and $p_2$ are the quark momenta flowing in and out of the
vertices) and of the quantity
\begin{equation}
X_0(p)=\frac{1}{a} \Bigg( i \sum_\mu \gamma_\mu \sin ap_\mu
+ 2r \sum_\mu \sin^2 \frac{ap_\mu}{2} -\rho \Bigg) .
\end{equation}
The quark-quark-gluon vertex in the overlap theory has the expression
\begin{eqnarray}
V_{1\mu}(p_1,p_2) &=& \rho \, \frac{1}{\omega(p_1) + \omega(p_2)} \times \\
&& \Bigg[ W_{1\mu} (p_1,p_2) -\frac{1}{\omega(p_1)\omega(p_2)} X_0(p_2) 
W_{1\mu}^\dagger (p_1,p_2) X_0(p_1) \Bigg] ,
\end{eqnarray}
and the quark-quark-gluon-gluon vertex is
\begin{eqnarray}
\lefteqn{V_{2\mu\nu}(p_1,p_2) = 
\delta_{\mu\nu} \, \rho \, \frac{1}{\omega(p_1) + \omega(p_2)} \times } 
\label{eq:v2} \\  
&&\qquad \Bigg[ W_{2\mu} (p_1,p_2) -\frac{1}{\omega(p_1)\omega(p_2)} X_0(p_2) 
W_{2\mu}^\dagger (p_1,p_2) X_0(p_1) \Bigg] \nonumber \\
&& +\frac{1}{2}\rho \, \frac{1}{\omega(p_1) 
+\omega(p_2)} \frac{1}{\omega(p_1) + \omega(k)} \frac{1}{\omega(k) + 
\omega(p_2)} \times \nonumber \\
&& \qquad \Bigg[X_0(p_2) W_{1\mu}^\dagger (p_2,k) W_{1\nu} (k,p_1) 
+W_{1\mu} (p_2,k) X_0^\dagger (k) W_{1\nu} (k,p_1) \nonumber \\
&& \qquad \qquad +W_{1\mu} (p_2,k) W_{1\nu}^\dagger (k,p_1) X_0 (p_1) 
\nonumber \\
&& \qquad -\frac{\omega(p_1)+\omega(k)+\omega(p_2)}{\omega(p_1)
\omega(k)\omega(p_2)}
X_0 (p_2) W_{1\mu}^\dagger (p_2,k) X_0 (k) W_{1\nu}^\dagger (k,p_1) X_0 (p_1)
\Bigg] . \nonumber
\end{eqnarray}

\section{Some analytic results}

We give here the analytic results for the leg tadpole and for the vertex of 
the scalar current, in the Feynman gauge for $r=1$. In order to be able to 
write them in a compact form it is convenient to introduce the abbreviations
\begin{equation}
M_\lambda = \cos \frac{k_\lambda}{2}, \quad 
N_\lambda = \sin \frac{k_\lambda}{2}, \quad
s_\lambda = \sin k_\lambda , \quad s^2 = \sum_\lambda s^2_\lambda,
\end{equation}
\begin{equation}
b=b(k)=2 \sum_\lambda \sin^2 \frac{k_\lambda}{2} -\rho,
\end{equation}
\begin{equation}
D = 2 \rho  \Big( \omega (k) + b(k) \Big) ,
\end{equation}
\begin{equation}
A = \frac{\omega^2(k)}{\rho^2}= 1-\frac{4}{\rho} \sum_\lambda \sin^2 
\frac{k_\lambda}{2} +\frac{1}{\rho^2} \Bigg( \sum_\lambda \sin^2 k_\lambda 
+ \left( 2 \sum_\lambda \sin^2 \frac{k_\lambda}{2} \right)^2 \Bigg)  . 
\end{equation}
The result for the 1-loop leg tadpole is then given by
\begin{eqnarray}
&& \frac{1}{2} g_0^2 \int \frac{d^4k}{2\pi^4} G(k) 
\Bigg( 1 - \frac{4}{\rho} \Bigg) \\
&& + g_0^2 \int \frac{d^4k}{2\pi^4} G(k) \frac{1}{\rho^2 (1+\sqrt{A})^2} 
\times \nonumber \\
&& \qquad \sum_\lambda \Bigg[ M_\lambda^2+N_\lambda^2 
+ \Bigg(1+\frac{1}{\sqrt{A}} \Bigg) \Bigg( - s_\mu^2 
+ b (M_\mu^2-N_\mu^2) \Bigg) \nonumber \\
&& \qquad \qquad + \frac{2+\sqrt{A}}{\rho\sqrt{A}} \Bigg(  
- b (M_\lambda^2-N_\lambda^2)  + s^2 \Bigg) \Bigg]  \nonumber \\
&& + g_0^2 \int \frac{d^4k}{2\pi^4} G^2(k) \frac{1}{\rho^2 (1+\sqrt{A})^2}
\times \nonumber \\
&& \qquad \sum_\lambda \Bigg[ -2 s_\mu^2 N_\lambda^2  
+ \Bigg(1+\frac{1}{\sqrt{A}} \Bigg) \Bigg( 2 s_\mu^2 
( b + 2 M_\mu^2 - M_\lambda^2 ) \Bigg) \Bigg] .  \nonumber 
\end{eqnarray}
The first term comes from the part of the $V_2$ vertex (\ref{eq:v2}) 
containing $W_2$ and $W_2^\dagger$, and its value for $\rho=1$ is 
\begin{equation}
g_0^2 \cdot \frac{Z_0}{2} \Bigg( 1 - \frac{4}{\rho} \Bigg) \Bigg|_{\rho=1}
= -\frac{g_0^2}{16\pi^2} \, 36.69915 ,
\end{equation}
while the 1-loop result for the whole leg tadpole is smaller:
\begin{equation}
-\frac{g_0^2}{16\pi^2} \, 23.35975 .
\end{equation}
Adding now the value $-\frac{g_0^2}{16\pi^2} \, 14.27088$ for the diagram 
e in Fig.~1, which is much harder to compute by hand and would be a very 
lengthy analytic expression anyway, gives the result 
$-\frac{g_0^2}{16\pi^2} \, 37.63063$ for the complete self-energy in the 
Feynman gauge for $\rho=1$.

The results for the 1-loop vertex diagram of the scalar operator (the sails 
are not present in this case) can be written in the form
\begin{equation}
g_0^2 \int \frac{d^4k}{2\pi^4} G(p-k) \frac{1}{(1+\sqrt{A})^2}
\Bigg[ \Bigg( -\frac{s^2}{D^2} + \frac{1}{4 \rho^2} \Bigg) \cdot X 
+ \frac{1}{\rho D} \cdot Y \Bigg] ,
\end{equation}
for small $p$, where
\begin{eqnarray}
X &=& \sum_\lambda \Bigg[ - (M_\lambda^2 - N_\lambda^2) 
+ \frac{1}{\rho \sqrt{A}} 2 b (M_\lambda^2 + N_\lambda^2) 
\nonumber \\
 && \qquad + \frac{1}{\rho^2 A} \Bigg( (s^2 -b^2) (M_\lambda^2 -N_\lambda^2) 
+2 b s^2 \Bigg) \Bigg] \nonumber \\
Y &=&  \sum_\lambda \Bigg[ s^2 + \frac{1}{\rho \sqrt{A}} 2s^2 
(M_\lambda^2 + N_\lambda^2) \nonumber \\
 && \qquad + \frac{1}{\rho^2 A} s^2 \Bigg( -2 b (M_\lambda^2 -N_\lambda^2) 
+ s^2 -b^2 \Bigg) \Bigg] .
\end{eqnarray}

We have checked that all these results obtained by hand correspond with the
outputs of the FORM programs.

\end{document}